\documentclass[12pt]{article}

\usepackage{amssymb}
\usepackage{amsmath}
\usepackage{latexsym}
\usepackage{yfonts}

\catcode`@=11
\renewcommand{\section}{\@startsection{section}{1}{0pt}{\medskipamount}
{\medskipamount}{\large\bf}} \numberwithin{equation}{section}
\catcode`@=12

\def\beq{\begin{eqnarray}}    
\def\eeq{\end{eqnarray}}      

\def\Box{\square}                       

\def\pa{\partial}                       

\def\={\ =\ }

\def\vp{\varepsilon}

\begin{document}

\begin{center}

{\Large\bf RG and BV-formalism }

\vspace{18mm}

{\large
Peter M. Lavrov$^{(a, b)}\footnote{E-mail:
lavrov@tspu.edu.ru}$\; }

\vspace{8mm}

\noindent  ${{}^{(a)}} ${\em
Tomsk State Pedagogical University,\\
Kievskaya St.\ 60, 634061 Tomsk, Russia}

\noindent  ${{}^{(b)}} ${\em
National Research Tomsk State  University,\\
Lenin Av.\ 36, 634050 Tomsk, Russia}

\vspace{20mm}

\begin{abstract}
\noindent
In present paper a quantization scheme proposed recently by Morris
(arXiv:1806.02206[hep-th]) is analyzed.
This method  is based on idea to combine
the renormalization group with the BV-formalism
in an unique  quantization procedure.
It is shown  that the BV-formalism and the new  method
should be considered as independent
approaches to quantization of gauge systems both provided by
global supersymmetry.

\end{abstract}

\end{center}

\vfill

\\

\newpage

\section{Introduction}

\noindent
At present the BRST symmetry \cite{brs1,t} is considered as a fundamental
principle of Modern Quantum
Field Theory allowing suitable quantum description of a given dynamical system
\cite{Weinberg,Green}. This principle is underlying the powerful quantization methods known
in covariant formalism as the Batalin-Vilkovisky (BV) method  \cite{BV,BV1} and
in canonical formulation as
the Batalin-Fradkin-Vilkovisky (BFV) approach \cite{BFV1,BFV2} (for recent
developments of these methods see \cite{BLT2015,BL2016,BL2017,BL2014,BL2018}).
Application of these methods to a given dynamical system
guarantees  gauge-independence of physical results
thanks  to the BRST symmetry.

The Gribov-Zwanziger theory \cite{Gribov,Zwanziger,Zwanziger1} and the
functional renormalization group approach \cite{Wet1,Wet2} belong to
a class of non-perturbative formulations of quantum theory of gauge fields with
violation of the BRST symmetry. In its turn the breakdown of the BRST symmetry
in both these cases leads to gauge dependence of effective action
even on-shell \cite{LLR,LL1,FRG-gauge,L,BLRNSh} making physical
interpretation of the results impossible.

Recently it has been proposed to combine methods of the BV-formalism
with the exact renormalization group (ERG)\cite{Morris1}. To do this
regularized versions of the antibracket and Delta-operator have been
introduced in a way dictated by main concepts of the renormalization
group with preserving (almost) all basic properties known from the
BV-formalism. In contrast with standard formulation of the
functional renormalization group approach \cite{Wet1,Wet2} the
regulators should be introduced  in a way that preserving gauge invariance
of regularized action
\cite{Morris1,Morris2,IIS}.

In present paper we analyze basic assumptions of new quantization
approach \cite{Morris1} in general gauge theories. As a result we
conclude that the BV-formalism and the ERG method \cite{Morris1}
should be considered as independent quantization schemes. It means
in particular that one needs to study basic properties of the new
quantization procedure and among them a gauge fixing procedure, a
gauge dependence problem,  existence of global supersymmetry and so
on. As in the case of BV-formalism the ERG can be provided with a
global supersymmetry (regularized BRST symmetry).

The paper is organized as follows. In section 2 a short presentation
of the BV-formalism is given. Basic ingredients and relations of
the ERG method are considered in section 3. In section 4 anticanonical
transformations in the ERG method and relations between the BV-formalism and the ERG method
are studied. In section 5 the gauge fixing procedure for general gauge theories
within the ERG method is introduced. In section 6 the existence of global supersymmetry
(regularized BRST symmetry) in the ERG method for the general gauge  and the Yang-Mills type
theories is proved. Finally, our conclusions and remarks
are presented in section 7.

In the paper the DeWitt's condensed notations are used \cite{DeWitt}.
We employ the notation $\varepsilon(A)$ for the Grassmann parity of
any quantity $A$.  The right and left functional derivatives with respect to
fields and antifields are marked by special symbols $"\leftarrow"$  and
$"\rightarrow"$ respectively.

\setcounter{section}{1}
\renewcommand{\theequation}{\thesection.\arabic{equation}}
\setcounter{equation}{0}

\section{BV-formalism in short}

\noindent The BV-formalism presents a powerful method of covariant
quantization of general gauge theories \cite{BV,BV1}. It is based on
a number of fundamental assumptions about the properties of the
systems in question. It is assumed that a given system of fields
$A^i$, $\vp(A^i)=\vp_i$ is described by an initial classical action
$S_0[A]$ being invariant under gauge transformations,
$\delta_{\xi}A^i=R^i_{\alpha}(A)\xi^{\alpha}$, where
$R^i_{\alpha}(A)$, $\vp(R^i_{\alpha})=\vp_i+\vp_{\alpha}$ are
generators of gauge transformations and
$\xi^{\alpha}=\xi^{\alpha}(x)$, $\vp(\xi^{\alpha})=\vp_{\alpha}$ are
arbitrary functions. In general algebra of gauge generators may be
(ir)reducible and (or) open and structure coefficients may depend on
fields. Taking into account the structure of gauge algebra one
defines a minimal antisymplectic space parameterized by fields
$\Phi^A_{min}$, $\vp(\Phi^A_{min})=\vp_A$ and antifields
$\Phi^{*\;min}_{A}$, $\vp(\Phi^{*\; min}_{A })=\vp_A+1$. For
irreducible gauge algebra the set $\Phi^A_{min}=\{A^i,C^{\alpha}\}$
includes initial fields $A^i$ and ghost fields $C^{\alpha}$,
$\vp(C^{\alpha})=\vp_{\alpha}+1$. In turn the set of corresponding
antifields takes the form $\Phi^{*\;min}_A=\{A^*_i,C^*_{\alpha}\}$.
For reducible theories the set $\Phi^A_{min}$ looks more complicate
and includes a pyramid of ghost for ghost fields and a pyramid of
auxiliary fields but here we are not going to details. In the
minimal antisymplectic space a solution,
$S_{min}=S_{min}[\Phi_{min}\Phi^{*\;min}]$,
 to the classical master equation, $(S_{min},S_{min})=0$,
 $S_{min}|_{\Phi^{*\;min}=0}=S_0[A]$, is constructed in the Taylor expansion
 with respect to ghost fields. Then full antisymplectic space of fields
 $\Phi=\{\Phi^A\}$ and antifields $\Phi^*=\{\Phi^*_A\}$ is introduced.
 For irreducible
 gauge algebra the explicit content of these sets are $\Phi^A=\{A^i,C^{\alpha},
 \bar{C}^{\alpha}, B^{\alpha}\}$,
 $\Phi^*_A=\{A^*_i,C^*_{\alpha}, \bar{C}^*_{\alpha}, B^*_{\alpha}\}$ where
 $\bar{C}^{\alpha}$ and $B^{\alpha}$ are antighost and auxiliary
 (Nakanishi-Lautrup) fields
 correspondingly. In full antisymplectic space the action $S=S[\Phi,\Phi^*]$
 constructed
 by the rule $S=S_{min}+\bar{C}^*_{\alpha}B^{\alpha}$ satisfies the quantum
 master equation
 $(1/2)(S,S)=i\hbar \Delta S$. This action is the initial object of
 the BV-formalism
 in construction of quantum description of a given gauge system.
 Making use a special type
 of anticanonical transformation with the help of Fermion
 functional $\Psi=\Psi[\Phi]$  the gauge-fixing functional
 $S_{\Psi}=S_{\Psi}[\Phi,\Phi^*]$ is introduced. The action $S_{\Psi}$
 satisfies the quantum
 master equation as well, $(1/2)(S_{\Psi},S_{\Psi})=i\hbar \Delta S_{\Psi}$.
 With the help of
 $S_{\Psi}$ the generating functional of Green functions, $Z[J,\Phi^*]$,
 in the form of functional integral over fields $\Phi$ is defined.
 Vacuum functional,
 $Z[\Phi^*]=Z[J=0,\Phi^*]$ obeys very important property of independence
 on specific choice
 of gauge-fixing functional $\Psi$ as a consequence
 that $S_{\Psi}$ satisfies the quantum
 master equation. In turn it means the gauge independence
 of physical quantities
 constructed in the BV-formalism due to the equivalence theorem \cite{KT}.

\section{Morris's construction}

\noindent
In the paper \cite{Morris1} a generalization of the antibracket and
the Delta-operator of the BV-formalism has been proposed. For any
two functionals $F=F[\Phi,\Phi^*]$ and $G=G[\Phi,\Phi^*]$ the new
antibracket is given by the rule
\beq
\label{AB}
(F,G)_{\Lambda}=\int dx
F\Big(\frac{\overleftarrow{\delta}}{\delta \Phi^A(x)} K_{\Lambda}(x)
\frac{\overrightarrow{\delta}}{\delta \Phi^*_A(x)}-
\frac{\overleftarrow{\delta}}{\delta \Phi^*_A(x)}K_{\Lambda}(x)
\frac{\overrightarrow{\delta}}{\delta \Phi^A(x)}\Big)G,
\eeq
or in
the DeWitt's condensed  notation
\beq
(F,G)_{\Lambda}=F\big(\overleftarrow{\pa}_{\Phi^A}\;
K_{\Lambda}\overrightarrow{\pa}_{\Phi^*_A}-
\overleftarrow{\pa}_{\Phi^*_A}\;K_{\Lambda}
\overrightarrow{\pa}_{\Phi^A}\big)\;\!G.
\eeq
Here $\Phi=\{\Phi^A\}, \vp(\Phi^A)=\vp_A$ and
$\Phi^*=\{\Phi^*_A\}, \vp(\Phi^*_A)=\vp_A+1$ are sets of fields and
antifields correspondingly and a regulator operator $K_{\Lambda}$ is introduced,
\beq
K_{\Lambda}(x)=K(\Box/\Lambda^2),\quad \Box=\pa_{\mu}\pa^{\mu},
\quad \vp(K_{\Lambda})=0, \eeq with the following properties:
$K_{\Lambda}(0)=1$ and  $K_{\Lambda}(x)\;\rightarrow 0$
for $x\rightarrow \infty$.

Taking into account that the integration by parts reads \beq
\label{ibp} \int dx f(x)K_{\Lambda}(x)g(x)=\int dx\;
g(x)K_{\Lambda}(x)f(x)(-1)^{\vp(f)\vp(g)}= \int
dx\;\big(K_{\Lambda}(x)f(x)\big)g(x) , \eeq one can check that  the
standard properties of the antibracket in the BV-formalism \cite{BV,BV1} hold
for the new antibracket (\ref{AB}) as well:

\noindent
(1) Grassmann parity
\begin{eqnarray}
\label{PropAntiBBV}
\varepsilon((F, G)_{\Lambda})=\varepsilon(F)+
\varepsilon(G)+1=\varepsilon((G, F)_{\Lambda})
\end{eqnarray}
(2) Generalized antisymmetry
\begin{eqnarray}
(F, G)_{\Lambda}=-(G, F)_{\Lambda}(-1)^{(\varepsilon(F)+1)(\varepsilon(G)+1)},
\end{eqnarray}
(3) Leibniz rule
\begin{eqnarray}
(F, GH)_{\Lambda}=(F, G)_{\Lambda}H+(F, H)_{\Lambda}
G(-1)^{\varepsilon(G)\varepsilon(H)},
\end{eqnarray}
(4) Generalized Jacobi identity
\begin{eqnarray}
((F, G)_{\Lambda}, H)_{\Lambda}(-1)^{(\varepsilon(F)+1)(\varepsilon(H)+1)}
+{\sf cycle} (F, G, H)\equiv 0.
\end{eqnarray}

The generalized Delta-operator has the form\footnote{As in the BV-formalism
the operator
$\Delta_{\Lambda}$ is ill-defined due to local nature of differential
operation entering in
(\ref{DLam}) and leading to the $\delta(0)$ problem. This problem can be solved
by using for example the dimensional regularization when $\delta(0)=0$.}
\beq
\label{DLam}
\Delta_{\Lambda}=\int dx\;(-1)^{\vp_A}
\frac{\overrightarrow{\delta}}{\delta \Phi^A(x)}
K_{\Lambda}(x)\frac{\overrightarrow{\delta}}{\delta \Phi^*_A(x)}=
(-1)^{\vp_A}\overrightarrow{\pa}_{\Phi^A}\;\!
K_{\Lambda}\;\!\overrightarrow{\pa}_{\Phi^*_A},\quad \vp(\Delta_{\Lambda})=1,
\eeq
and obeys the nilpotency property
\beq
\Delta^2_{\Lambda}=0.
\eeq

Action of the generalized Delta-operator on the antibracket (\ref{AB})
takes the standard form in the BV-formalism
\beq
\label{dp}
\Delta_{\Lambda}(F,G)= (\Delta_{\Lambda}F,G)_{\Lambda}-
(F,\Delta_{\Lambda}G)_{\Lambda}(-1)^{\vp(F)}.
\eeq The same
statement is valid for action of the generalized Delta-operator on
product of two functionals. The result reads
\beq
\label{dp1}
\Delta_{\Lambda}\big(F \cdot G\big)=\big(\Delta_{\Lambda}F\big)\cdot
G + F\cdot\big(\Delta_{\Lambda}
G\big)(-1)^{\vp(F)}+(F,G)_{\Lambda}(-1)^{\vp(F)}. \eeq
In deriving (\ref{dp}) and (\ref{dp1}) the integration by parts (\ref{ibp})
was intensively used.


For a given gauge theory with initial classical action $S_0[A]$
having properties described in Sec. 2  it is assumed that there
exists a possibility to construct a regularized action
$S_{0\Lambda}[A]$ which will be  invariant under (regularized) gauge
transformations. It is exactly the case of paper \cite{IIS} where
regularization of kinematic part of full action is presented in
explicit form and existence of regularized part of interaction
assumed. In paper \cite{Morris1} the regularization of
interaction terms of full action in the case of non-Abelian gauge
theories is given perturbatively.
Then it is expected that all the
consequences of the BV formalism can be applied. This would be so if
the new formalism were equivalent to the BV formalism. We are going to
prove non-equivalence of the BV-formalism and the new quantization method.

\section{Anticanonical transformations}
\noindent Anticanonical transformation in the BV-formalism preserves
structure of any relation involving the antibracket and the
Delta-operator. Relations listed in Section 3 confirmed that.

But there exist an essential difference between these two
approaches. It is related with canonical relations in the
BV-formalism,
\beq
\label{CR-BV}
(\Phi^A,\Phi^*_B)=\delta^A_{\;B},
\eeq
and relations with the regularized antibracket,
\beq
\label{rcr}
(\Phi^A,\Phi^*_B)_{\Lambda}=K_{\Lambda}\delta^A_{\;B}.
\eeq
The relations (\ref{rcr}) can be rewritten in the form
\beq
(\Phi^A,K^{-1}_{\Lambda}\Phi^*_B)_{\Lambda}=
(K^{-1}_{\Lambda}\Phi^A,\Phi^*_B)_{\Lambda}=\delta^A_{\;B}.
\eeq
When $K_{\Lambda}\neq 1$ there is no anticanonical transformation
reproducing the relations
\beq
\label{rcr1} \Phi^{'A}=\Phi^A,\quad
\Phi^*_A=K_{\Lambda}\Phi^{*'}_A\;\quad {\rm  or}\;\quad
\Phi^{'A}=K^{-1}_{\Lambda}\Phi^A, \quad \Phi^*_A=\Phi^{*'}_A .
\eeq
Indeed, let $F=F[\Phi,\Phi^{*'}]$, $\vp(F)=1$ be generator of
anticanonical transformation,
\beq
\Phi^{'A}=\overrightarrow{\pa}_{\Phi^{*'}_A}F[\Phi,\Phi^{*'}],\quad
\Phi^*_A=F[\Phi,\Phi^{*'}]\overleftarrow{\pa}_{\Phi^A}.
\eeq
Then
from (\ref{rcr1}) we have, in particular,
\beq
\overrightarrow{\pa}_{\Phi^{*'}_A}F[\Phi,\Phi^{*'}]=\Phi^A
\eeq
and
therefore
\beq F[\Phi,\Phi^{*'}]=\Phi^{*'}_A\Phi^A+X[\Phi], \quad
\vp(\Psi)=1,
\eeq
with some odd functional $X=X[\Phi]$. The second
relation in (\ref{rcr1}) allows us to specify the functional $X[\Phi]$,
\beq
(K_{\Lambda}-1)\Phi^{*'}_A=X[\Phi]\overleftarrow{\pa}_{\Phi^A},
\eeq
with the results $X[\Phi]={\rm const}$, $K_{\Lambda}=1$.
Because anticanonical transformations (automorphism of CME)
\cite{VLT} and anticanonical master-transformations (automorphism of
QME) \cite{LT-82,LT-85,BLT2015} rule all scheme of the BV-formalism
we conclude that the ERG method is independent quantization
procedure for general gauge theories.

The regularized antibracket is not invariant under anticanonical transformations accepted
in the BV-formalism, where anticanonical transformations play very important role in solving
all the principal problems concerning the gauge fixing procedure,
the gauge invariant renormalization, the gauge dependence problem and so on.
As an independent approach the ERG method based on using
the regularized quantum master equation formulated in terms of regularized antibracket
and regularized Delta-operator requires first of all to study the invariance properties
of the regularized antibracket. Now we have to consider the relation
\beq
(\Phi^A(x),\Phi^*_B(y))=\delta^A_{\;B}K_{\Lambda}(x)\delta(x-y)
\eeq
 as the
basic one in the ERG method. It is clear that the identical transformation leaving
the regularized antibracket invariant is described by the functional
$X_{0\Lambda}[\Phi,\Phi^{*'}]$,
\beq
X_{0\Lambda}[\Phi,\Phi^{*'}]=\int dx \Phi^{*'}_A(x)K_{\Lambda}(x)\Phi^A(x).
\eeq
We consider now transformations of variables which infinitesimally differ of the identical
ones and describe by the functional
$F[\Phi,\Phi^{*'}]_{\Lambda}=X_{0\Lambda}[\Phi,\Phi^{*'}]+\varepsilon[\Phi,\Phi^{*'}]$,
so that
\beq
\label{Tr1}
\Phi^{'A}(x)&=&\overrightarrow{\pa}_{\Phi^{*'}_A(x)}F_{\Lambda}[\Phi,\Phi^{*'}]=
K_{\Lambda}(x)\Phi^A(x)+\overrightarrow{\pa}_{\Phi^{*'}_A(x)}
\varepsilon[\Phi,\Phi^{*'}],\\
\label{Tr2}
\Phi^*_A(x)&=&F_{\Lambda}[\Phi,\Phi^{*'}]\overleftarrow{\pa}_{\Phi^A(x)}=
\Phi^{*'}_A(x)K_{\Lambda}(x)+
\varepsilon[\Phi,\Phi^{*'}]\overleftarrow{\pa}_{\Phi^A(x)}.
\eeq
In the first order in $\varepsilon[\Phi,\Phi^{*'}]$ we obtain
\beq
&&\qquad\qquad\quad(\Phi^{'A}(x),\Phi^{*'}_B(y))_{\Lambda}=
\delta^A_{\;B}K_{\Lambda}(x)\delta(x-y)+\\
\nonumber
&&+\overrightarrow{\pa}_{\Phi^{*}_A(x)}
\varepsilon[\Phi,\Phi^{*}]\overleftarrow{\pa}_{\Phi^B(y)}-
K_{\Lambda}(x)K_{\Lambda}(x)\overrightarrow{\pa}_{\Phi^{*}_A(x)}
\varepsilon[\Phi,\Phi^{*}]\overleftarrow{\pa}_{\Phi^B(y)}K^{-1}_{\Lambda}(y).
\eeq
In general, the regularized antibracket is not invariant under the transformations
(\ref{Tr1}), (\ref{Tr2}). Nevertheless, there exist the two cases
of transformations  (\ref{Tr1}), (\ref{Tr2})
preserving the form of regularized antibracket. They correspond to the following
choice of functional $\varepsilon[\Phi,\Phi^{*}]$,
\beq
\varepsilon[\Phi,\Phi^{*}]=\Psi[\Phi]\;\quad {\rm or}\quad
\varepsilon[\Phi,\Phi^{*}]=X[\Phi^*]
\eeq
for arbitrary odd functionals $\Psi[\Phi]$, $X[\Phi^*]$. We can refer to these cases as
reduced anticanonical transformations in the ERG method.


\section{Gauge fixing procedure}

Now we are in position to describe the gauge fixing procedure
in the new quantization approach in general case of any initial system of gauge fields
$A^i$ with an action $S_0[A]$ which is invariant under the gauge transformations
$\delta A^i=R^i_{\alpha}(A)\xi^{\alpha}$.
Main idea of the EFG method is to save
the gauge invariance of regularized action $S_{0\Lambda}[A]$ corresponding to $S_0[A]$,
\beq
\label{F1}
S_{0\Lambda}[A]\overleftarrow{\pa}_{\!\!A^i}R^i_{\Lambda\alpha}(A)\xi^{\alpha}=0.
\eeq
The regularized action $S_{0\Lambda}[A]$ and regularized gauge generators
$R^i_{\Lambda\alpha}(A)$ satisfy the properties
\beq
\label{F2}
\lim_{\Lambda\rightarrow 0}S_{0\Lambda}[A]=S_0[A],\quad
\lim_{\Lambda\rightarrow 0}R^i_{\Lambda\alpha}(A)=R^i_{\alpha}(A).
\eeq
Requirement of gauge invariance of the regularized action (\ref{F1}) differs from usually
accepted regularization of kinematic part of full action
in the standard FRG approach \cite{Wet1,Wet2} which leads to breakdown
of gauge symmetry and causes
the gauge dependence of effective average action even on-shell \cite{FRG-gauge,L}.

Now  let $S_{\Lambda}=S_{\Lambda}[\Phi,\Phi^*]$
be an action satisfying the regularized quantum master equation
\begin{eqnarray}
\label{F3}
\frac {1}{2} (S_{\Lambda},S_{\Lambda})_{\Lambda}=
i\hbar{\Delta}_{\Lambda}S_{\Lambda}\;\leftrightarrow\;
\Delta_{\Lambda}\exp\Big\{\frac{i}{\hbar}S_{\Lambda}\Big\}=0,
\end{eqnarray}
and the boundary condition
\begin{eqnarray}
\label{F4}
S_{\Lambda}\big|_{\Phi^* = \hbar = 0}= S_{0\Lambda}[A].
\end{eqnarray}

For any odd functional $\Psi=\Psi[\Phi]$ we derive the relation
\beq
\label{F5}
[\Delta_{\Lambda}, \Psi]= -\int dx
\big(\Psi[\Phi]\overleftarrow{\pa}_{\!\!\Phi^A(x)}\big)K_{\Lambda}(x)
\overrightarrow{\pa}_{\!\!\Phi^*_A(x)},
\eeq
which allows to state that
\beq
\label{F6}
\exp\big\{-[\Delta_{\Lambda}, \Psi]\big\}W[\Phi,\Phi^*]=
W[\Phi,\Phi^*+\Psi[\Phi]\overleftarrow{\pa}_{\!\!\Phi}K_{\Lambda}],
\eeq
where $W[\Phi,\Psi^*]$ is arbitrary functional. In particular, we can construct
the action $S_{\Lambda\Psi}=S_{\Lambda\Psi}[\Phi,\Phi^*]$ as
\beq
\label{F7}
S_{\Lambda\Psi}[\Phi,\Phi^*]=
\exp\big\{-[\Delta_{\Lambda}, \Psi]\big\}S_{\Lambda}[\Phi,\Phi^*]=
S_{\Lambda}[\Phi,\Phi^*+\Psi[\Phi]\overleftarrow{\pa}_{\!\!\Phi}K_{\Lambda}],
\eeq
where $S_{\Lambda}[\Phi,\Phi^*]$ is a solution to equations (\ref{F3}), (\ref{F4}). The
 functional (\ref{F7}) satisfies the quantum master equation (\ref{F3}) as well. Indeed,
 let us compute the commutator of regularized Delta-operator $\Delta_{\Lambda}$ with
 $[\Delta_{\Lambda}, \Psi]$. The result reads
\beq
\label{F8}
[\Delta_{\Lambda},[\Delta_{\Lambda}, \Psi]]=-
\int dxdy (-1)^{\varepsilon_A(\varepsilon_B+1)}
\big(\Psi[\Phi]\overleftarrow{\pa}_{\!\!\Phi^B(y)}\overleftarrow{\pa}_{\!\!\Phi^A(x)}\big)
K_{\Lambda}(y)\overrightarrow{\pa}_{\!\!\Phi^*_B(y)}
K_{\Lambda}(x)\overrightarrow{\pa}_{\!\!\Phi^*_A(x)}.
\eeq
Due to the symmetry properties of integrand in (\ref{F8}) we conclude that
\beq
\label{F9}
[\Delta_{\Lambda},[\Delta_{\Lambda}, \Psi]]=0.
\eeq
From Eqs. (\ref{F3}), (\ref{F7}), (\ref{F9}) it follows
\beq
\nonumber
&&0=\exp\big\{-[\Delta_{\Lambda}, \Psi]\big\}\Delta_{\Lambda}
\exp\Big\{\frac{i}{\hbar}S_{\Lambda}\Big\}=
\Delta_{\Lambda}\exp\big\{-[\Delta_{\Lambda}, \Psi]\big\}
\exp\Big\{\frac{i}{\hbar}S_{\Lambda}\Big\}=\\
\label{F10}
&&\quad=
\Delta_{\Lambda}\exp\Big\{\frac{i}{\hbar}S_{\Lambda\Psi}\Big\}=0.
\eeq
We consider the relation (\ref{F7}) as gauge fixing procedure in the ERG method.
In the limit $\Lambda\rightarrow 0$ the procedure described coincides with the gauge
fixing procedure in the BV-formalism where $\Psi[\Phi]$ is the gauge fixing functional.

\section{Global supersymmetry}
\noindent
Now we are going to prove that  there
exists a possibility to provide the ERG method with a global supersymmetry
similarly to the BV-formalism. Starting point is the generating functional of Green
function
\beq
\label{S1}
Z_{\Lambda}[J,\Phi^*]=\int D\Phi\exp\Big\{\frac{i}{\hbar}
\big(S_{\Lambda\Psi}[\Phi,\Phi^*]+J_A\Phi^A\big)\Big\},
\eeq
where $J_A$ $( \varepsilon(J_A)=\varepsilon(\Phi^A)=\varepsilon_A)$ are external sources
to fields $\Phi^A$. To discuss the existence of global supersymmetry within the ERG method
we consider the vacuum functional $Z_{\Lambda}=Z_{\Lambda}[J=0,\Phi^*=0]$,
\beq
\label{S2}
Z_{\Lambda}=\int d\Phi \;\exp\Big\{\frac{i}{\hbar}
S_{\Lambda}[\Phi,\Psi[\Phi]\overleftarrow{\pa}_{\!\!\Phi}K_{\Lambda} ]\Big\},
\eeq
It is convenient to present (\ref{S2}) in the form
\beq
\label{S3}
Z_{\Lambda}=\int d\Phi \; d\Phi^*\; d\lambda\;\exp\Big\{\frac{i}{\hbar}
\big(S_{\Lambda}[\Phi,\Phi^*]
+(\Phi^*_A -\Psi[\Phi]\overleftarrow{\pa}_{\!\!\!\Phi^A}K_{\Lambda})\lambda^A\big)\Big\},
\eeq
with the help of auxiliary fields
$\lambda^A,\;\;\varepsilon(\lambda^A) = \varepsilon_A + 1$.

Then the integrand in (\ref{S3}) is
invariant under the following global supertransformations:
\begin{eqnarray}
\label{S4}
\delta_{\Lambda}\Phi^A = K_{\Lambda}\lambda^A\mu,\quad \delta_{\Lambda}\Phi^*_A = \mu
K_{\Lambda}\big(S_{\Lambda}[\Phi,\Phi^*]\overleftarrow{\pa}_{\!\!\Phi^A}\big),
\quad\delta_{\Lambda}\lambda^A = 0,
\end{eqnarray}
where $\mu$ is a constant anticommuting parameter. In deriving this
result we take into account the following facts: a) the Jacobian is equal to
\beq
\label{S5}
J=\exp\{\Delta_{\Lambda}S_{\Lambda}\},
\eeq
b) the functional $S_{\Lambda}$ satisfies
the regularized quantum master equation (\ref{F3}),
c) the equality holds
\beq
\label{S6}
\int dxdy (-1)^{\varepsilon_A(\varepsilon_B+1)}
\big(\Psi[\Phi]\overleftarrow{\pa}_{\!\!\Phi^B(y)}\overleftarrow{\pa}_{\!\!\Phi^A(x)}\big)
K_{\Lambda}(y)\lambda^B(y)
K_{\Lambda}(x)\lambda^A(x)=0
\eeq
due to the symmetry properties in the integrand (\ref{S6}).

The transformations (\ref{S4}) represent the global supersymmetry for the ERG approach
in the space of variables $\Phi,\;\Phi^*,\;\lambda$. They  may be
named as regularized BRST transformations.
Notice that in general the regularized BRST symmetry is not symmetry of
some action in full agreement with situation
in the BV-formalism. There is another property of transformations (\ref{S4})
similarly to the BRST transformations in the BV-formalism, namely
they do not depend on choice of gauge fixing condition.
It is very important to realize that the existence of this symmetry is
the consequence of the fact that the bosonic functional $S_{\Lambda}$
satisfies the regularized quantum master equation (\ref{F3}).

In the case when the regularized action $S_{0\Lambda}[A]$  belongs to
Yang-Mills type of gauge theories,
the action $S_{\Lambda\Psi}[\Phi,\Phi^*]$  can be constructed in explicit form.
To this end we assume gauge invariance of the action $S_{0\Lambda}[A]$,
\beq
\label{S7}
S_{0\Lambda, i}[A]R^i_{\Lambda \alpha}(A)=0.
\eeq
where the gauge generators $R^i_{\Lambda \alpha}(A)$ form a linear independent set
in the index $\alpha$ and satisfy the gauge algebra
\beq
\label{S8}
R^i_{\Lambda \alpha,j}(A)R^j_{\Lambda \beta}(A)-
(-1)^{\varepsilon_{\alpha}\varepsilon_{\beta}}
R^i_{\Lambda \beta,j}(A)R^j_{\Lambda \alpha}(A)=
-R^i_{\Lambda\gamma}(A)F^{\gamma}_{\Lambda\alpha\beta}.
\eeq
Here $F^{\gamma}_{\Lambda\alpha\beta}$ are structure coefficients which do not depend on
fields $A^i$  and the notation $G_{,i}=G\overleftarrow{\pa}_{\!\!A^i}$ is used. Now the action
$S_{\Lambda\Psi}[\Phi,\Phi^*]$ is constructed by the Faddeev-Popov rules \cite{FP}
and has the form
\beq
\label{S9}
S_{\Lambda\Psi}[\Phi,\Phi^*]=S_{0\Lambda}[A]+
\Psi[\Phi]\overleftarrow{\pa}_{\!\!\Phi^A}R^A_{\Lambda}(\Phi)+
\Phi^*_AR^A_{\Lambda}(\Phi),
\eeq
where $\Phi^A=(A^i, B^{\alpha}, C^{\alpha}, {\bar C}^{\alpha})$ are fields appearing in
the Faddeev-Popov method, and
\beq
\label{S10}
R^A_{\Lambda}(\Phi)=\big(R^i_{\Lambda \alpha}(A)C^{\alpha}, 0,
-(1/2)(-1)^{\varepsilon_{\beta}}
F^{\alpha}_{\Lambda\beta\gamma}C^{\gamma}C^{\beta},
(-1)^{\varepsilon_{\alpha}}B^{\alpha}\big)
\eeq
are  generators of the regularized BRST transformations
\beq
\label{S11}
\delta_{B\Lambda}\Phi^A=R^A_{\Lambda}(\Phi)\mu, \quad \mu={\rm const}, \quad
\varepsilon(\mu)=1.
\eeq
Due to the nilpotency of transformations (\ref{S11}),
\beq
\label{S12}
R^A_{\Lambda}(\Phi)\overleftarrow{\pa}_{\!\!\Phi^B}R^B_{\Lambda}(\Phi)=0,
\eeq
the action (\ref{S9}) is invariant under the regularized
BRST transformations (\ref{S11}).

\section{Discussion}

\noindent
In the paper we have analyzed basic assumptions of new approach for quantization of gauge
systems to combine attractive features of the BV-formalism with main idea of
the renormalization group \cite{Morris1}. We have confirmed basic algebraic
properties and the Jacobi identity for regularized antibracket and Delta-operator
introduced in \cite{Morris1} which are similarly to corresponding
relations in the BV-formalism. Nevertheless the BV-formalism \cite{BV,BV1}
and the new method \cite{Morris1}
should be considered as independent approaches to quantization
of gauge systems because
there does not exists an anticanonical transformations allowing
to connect the canonical
relations between fields and antifields (\ref{CR-BV}) in the BV-formalism and
corresponding relations in Morris's approach (\ref{rcr}).
We have found transformations of variables of antisymplectic space leading
to the invariance of  the regularized antibracket.
These transformations have been  called as
the reduced anticanonical transformations.

Notice that the cornerstone  of the new quantization approach is
construction of regularized action for an initial gauge theory. It
is not trivial problem because  it is required for the regularized  action
to be gauge invariant. In this point there is essential difference between the
standard FRG approach \cite{Wet1,Wet2} and the new ERG method
\cite{IIS,Morris1,Morris2}. Regularization of kinematic part of full
classical action accepted in the FRG violates the gauge invariance
that causes the gauge dependence problem of effective average action
even on-shell \cite{FRG-gauge,L}. The regularization of initial gauge invariant
action in the ERG method  needs in special efforts \cite{IIS,Morris1,Morris2}.
We want to illustrate problems arising in this way using pure
Yang-Mills theory with action \beq S_{0}[A] &=&-\frac{1}{4}\int dx
F_{\mu \nu }^{a}(x)F_{\mu \nu }^{a}(x)=-\frac{1}{4}
F_{\mu \nu }^{a}F_{\mu \nu }^{a},   \label{YM} \\
F_{\mu \nu }^{a }(x) &=&\partial _{\mu }A_{\nu }^{a }(x)-\partial
_{\nu }A_{\mu }^{\alpha }(x)+f^{abc }A_{\mu }^{b}(x)A_{\nu }^{c
}(x),
\label{FYM}
\eeq
where $f^{abc}$ are structure coefficients of
the $SU(N)$ Lie group satisfying the Jacobi identity,
\beq
f^{abc}f^{cde}+f^{aec}f^{cbd}+f^{adc}f^{ceb}\equiv 0.
\eeq
Let us try to introduce the regularization in the form preserving
the geometric description in terms of field strength $F_{\mu \nu }^{a}$.
It can be done for example in the form,
\beq
\label{RYM} S_{0\Lambda}[A]=-\frac{1}{4}\int
dx\;F^a_{\mu\nu}(x)K^{-1}_{\Lambda}(x)F^a_{\mu\nu}(x)=
-\frac{1}{4}\;F^a_{\mu\nu}K^{-1}_{\Lambda}F^a_{\mu\nu}.
\eeq
The
action (\ref{RYM}) is invariant under the following gauge
transformation
\beq
\label{vF}
\delta_{\xi\Lambda}F^a_{\mu\nu}=f^{abc}K_{\Lambda}F^b_{\mu\nu}\xi^c,\quad
\delta_{\xi\Lambda}S_{0\Lambda}[A]=0.
\eeq
From (\ref{FYM}) it
follows the presentation of this variation in terms of field
variations
\beq
\label{vFYM}
\delta_{\xi\Lambda}F^a_{\mu\nu}=
\pa_{\mu}\delta_{\xi\Lambda}A^{a}_{\nu}-
\pa_{\nu}\delta_{\xi\Lambda}A^{a}_{\mu}+f^{abc}
\big[\delta_{\xi\Lambda}A^{b}_{\mu}\;\!A^{c}_{\nu}+
A^{b}_{\mu}\;\!\delta_{\xi\Lambda}A^{c}_{\nu}\big].
\eeq
In the
Yang-Mills theory ($K_{\Lambda}=1$) the gauge transformations of the
field strength $F^a_{\mu\nu}$,
$\delta_{\xi}F^a_{\mu\nu}=f^{abc}F^b_{\mu\nu}\xi^c$, can be
rewritten in terms of gauge transformations of fields $A^a_{\mu}$,
$\delta_{\xi}A^a_{\mu}=D^{ab}_{\mu}(A)\xi^b$.
Let us try to present
the variation (\ref{vF}) in the form (\ref{vFYM}).
The result reads
\beq
\nonumber
\delta_{\xi\Lambda}F^a_{\mu\nu}&=&
\pa_{\mu}\big(K_{\Lambda}D^{ab}_{\nu}(A)\xi^b\big)-
\pa_{\nu}\big(K_{\Lambda}D^{ab}_{\mu}(A)\xi^b\big)+\\
\nonumber && \label{nGT}
 +f^{abc}\big[\big(K_{\Lambda}D^{bd}_{\mu}(A)\xi^d\big)A^c_{\nu}+
 \big(K_{\Lambda}D^{bd}_{\nu}(A)\xi^d\big)A^c_{\mu}\big]+\\
 &&+ f^{abc}[K_{\Lambda}, A^b_{\mu}]D_{\nu}^{cd}(A)\xi^d+
 f^{abc}[K_{\Lambda}, A^c_{\nu}]D_{\nu}^{bd}(A)\xi^d.
\eeq
For Abelian Lie group $f^{abc}=0$ one can formulate the gauge
invariance of the regularized initial action in terms   of gauge
transformations of fields $A_{\mu}$ as
$\delta_{\xi\Lambda}A_{\mu}=K_{\Lambda}\pa_{\mu}\xi$ and after that
to apply the  BV-formalism to construct suitable quantum description
of renormalization group respecting BRST symmetry.
In particular the BRST transformations in the sector of fields
$A_{\mu}$ are described by the relations
$\delta_{B\Lambda}A_{\mu}=K_{\Lambda}\pa_{\mu}C$.

In general from (\ref{nGT}) it follows that gauge invariance of the
action (\ref{RYM}) cannot be expressed in terms of gauge
transformations of fields $A^a_{\mu}$. In particular the gauge
transformations of fields $A^a_{\mu}$,
\beq
\delta_{\xi\Lambda}A^a_{\mu}=K_{\Lambda}D^{ab}_{\nu}(A)\xi^b,
\eeq
do not present symmetry transformations of the regularized action
(\ref{RYM}). It means that  the regularized version of initial
classical action  proposed here does not give a possibility to apply
the BV-formalism.
From point of view of the new ERG method this feature forbids
to use the regularization
procedure proposed here.

We have proved that the ERG method  can be provided with
reduced anticanonical transformations preserving the regularized antibracket
and with global supersymmetry (regularized BRST symmetry) if
assumptions listed in Secs. 5, 6 are fulfilled.
Main of them are the regularized initial gauge-invariant action (\ref{F1}),
the existence of solutions to quantum master equation (\ref{F3})
with boundary condition (\ref{F4}) and the gauge fixing procedure proposed (\ref{F7}).

\section*{Acknowledgments}
\noindent
The author thanks Igor Tyutin for useful discussions. The author is grateful
to anonymous referee for constructive
criticism which helped to extend results concerning  basic properties of the ERG method.
The work  is supported  by the RFBR grant 18-02-00153.


\begin {thebibliography}{99}
\addtolength{\itemsep}{-8pt}

\bibitem{brs1}
C. Becchi, A. Rouet, R. Stora, {\it The abelian Higgs Kibble
Model, unitarity of the $S$-operator}, Phys. Lett. B {\bf 52} (1974)
344.

\bibitem{t}
I.V. Tyutin, {\it Gauge invariance in field theory and statistical
physics in operator formalism}, Lebedev Institute preprint  No.  39
(1975), arXiv:0812.0580 [hep-th].

\bibitem{Weinberg}
S. Weinberg, {\it The quantum theory of fields, v.II},
Cambridge University Press, Cambridge, 1996.


\bibitem{Green}
M.B. Green, J.H. Schwarz, E. Witten,
{\it Superstring theory}, Cambridge University Press, Cambridge, 1988.

\bibitem{BV} I.A. Batalin, G.A. Vilkovisky, \textit{Gauge algebra and
quantization}, Phys. Lett. \textbf{B102} (1981) 27.

\bibitem{BV1} I.A. Batalin, G.A. Vilkovisky, \textit{Quantization of gauge
theories with linearly dependent generators}, Phys. Rev.
\textbf{D28} (1983)
2567.

\bibitem{BFV1}
E.S. Fradkin, G.A. Vilkovisky, {\it Quantization of relativistic
systems with constraints},
Phys. Lett. \textbf{B55} (1975) 224.

\bibitem{BFV2}
I.A. Batalin, G.A. Vilkovisky, {\it Relativistic S-matrix of dynamical systems
with boson and fermion constraints},
Phys. Lett. \textbf{B69} (1977) 309.

\bibitem{BLT2015}
I.A. Batalin, P.M. Lavrov, I.V. Tyutin,
{\it Finite anticanonical transformations in field-antifield formalism}.
Eur. Phys. J. \textbf{C75} (2015) 270.

\bibitem{BL2016}
I.A. Batalin, P.M. Lavrov,
{\it Closed description of arbitrariness in resolving quantum master equation},
Phys. Lett. {\bf  B758} (2016) 54.

\bibitem{BL2017}
I.A. Batalin, P.M. Lavrov,
{\it Superfield generating equation of field-antifield formalism as a
hyper-gauge theory}, Eur. Phys. J. {\bf C77} (2017) 121.

\bibitem{BL2014}
I.A. Batalin, P.M. Lavrov, I.V. Tyutin, 
{\it A systematic study of finite BRST-BFV transformations in generalized
Hamiltonian formalism},
Int. J. Mod. Phys. {\bf A29} (2014) 1450127.


\bibitem{BL2018}
I.A. Batalin, P.M. Lavrov,
{\it General conversion method for constrained systems},
Phys. Lett. {\bf  B787} (2018) 89.


\bibitem{Gribov}
V.N. Gribov, {\it Quantization of nonabelian gauge theories},
Nucl. Phys. {\bf B139} (1978) 1.

\bibitem{Zwanziger}
D. Zwanziger, {\it Action from Gribov horizon},
Nucl. Phys. {\bf B321} (1989) 591.

\bibitem{Zwanziger1}
D. Zwanziger, {\it Local and renormalizable action from the Gribov horizon},
Nucl. Phys. {\bf B323} (1989) 513.


\bibitem{Wet1}
C. Wetterich, {\it Average action and the renormalization group equation},
Nucl. Phys.  {\bf B352} (1991) 529.

\bibitem{Wet2}
C. Wetterich, {\it Exact evolution equation for the effective potential},
Phys. Lett. {\bf B301} (1993) 90.


\bibitem{LLR}
P. Lavrov, O. Lechtenfeld, A. Reshetnyak, {\it Is soft breaking of
BRST symmetry consistent?}
JHEP {\bf 1110} (2011) 043.

\bibitem{LL1}
P.M.  Lavrov, O. Lechtenfeld, {\it Gribov horizon beyond the Landau gauge},
Phys. Lett. {\bf B725} (2013) 386.

\bibitem{FRG-gauge} P.M.~Lavrov, I.L.~Shapiro,
{\it On the Functional Renormalization Group approach for
Yang-Mills fields,}
JHEP {\bf 1306} (2013) 086.

\bibitem{L}
P.M. Lavrov,
{\it Gauge (in)dependence and background field formalism}
Phys. Lett. {\bf B791} (2019) 293.

\bibitem{BLRNSh}
V.F. Barra, P.M. Lavrov, E.A. dos Reis, T. de Paula Netto, I.L. Shapiro,
{\it Functional renormalization group approach and gauge dependence in gravity
theories}, arXiv:1910.06068[hep-th].


\bibitem{Morris1}
T.R. Morris, {\it Quantum gravity, renormalizability and
diffeomorphism invariance}, SciPost Phys. {\bf 5} (2018) 040,
arXiv:1806.02206[hep-th].

\bibitem{Morris2}
Y. Igarashi, K. Itoh, T.R. Morris, {\it BRST in the exact renormalization
group}, Prog. Theor. Exp. Phys. (2019),
arXiv:1904.08231[hep-th].

\bibitem{IIS}
Y. Igarashi, K. Itoh, H. Sonoda,
{\it Realization  of  Symmetry  in  the  ERG Approach
to Quantum Field Theory}, Prog.
Theor. Phys. Suppl. {\bf 181} (2009) 1.

\bibitem{DeWitt}
B.S. DeWitt, \textit{Dynamical theory of groups and fields},
(Gordon and Breach, 1965).

\bibitem{KT}
R.E. Kallosh, I.V.  Tyutin,
{\it The equivalence theorem and gauge invariance in
renormalizable theories},
Sov. J. Nucl. Phys.
{\bf 17} (1973) 98.

\bibitem{VLT}
B.L. Voronov, P.M. Lavrov, I.V. Tyutin,
{\it Canonical transformations and gauge dependence
in general gauge theories},
Yad. Fiz.
{\bf 36} (1982) 498.

\bibitem{LT-82}
P.M. Lavrov, I.V. Tyutin,
{\it Gauge theories of general form},
Sov. Phys. J. {\bf 25} (1982) 639.

\bibitem{LT-85}
P.M. Lavrov, I.V. Tyutin,
{\it Effective action in general gauge theories},
Yad. Fiz.  {\bf 41} (1985) 1658.


\bibitem{FP}
L.D. Faddeev, V.N. Popov,
{\it Feynman diagrams for the Yang-Mills field},
Phys. Lett. {\bf B25} (1967) 29.

\end{thebibliography}
\end{document}